# DYNAMIC NETWORK OF CONCEPTS FROM WEB-PUBLICATIONS

*Lande D.V. (dwl@visti.net), IC «ELVISTI», NTUU «KPI»*
*Snarskii A.A. (asnarskii@gmail.com), NTUU «KPI»*

The network, the nodes of which are concepts (people's names, companies' names, etc.), extracted from web-publications, is considered. A working algorithm of extracting such concepts is presented. Edges of the network under consideration refer to the reference frequency which depends on the fact how many times the concepts, which correspond to the nodes, are mentioned in the same documents. Web-documents being published within a period of time together form an information flow, which defines the dynamics of the network studied. The phenomenon of its structure stability, when the number of web-publications, constituting its formation bases, increases, is discussed.

The analysis of complex networks having a social nature is a topic of present interest in the research. Recently, a separate branch of discrete mathematics, which is called the theory of complex networks, has been formed; it studies network characteristics taking into account not only their topology, but also the distribution of reference frequency of individual nodes [1]. Today this is a very actual theory in identifying and visualizing various communities, their internal correlations. A fast development of the Internet content made a great impetus on the development of theoretical and applied issues of the theory of complex networks. This research is devoted to the analysis of the network relations of the concepts (people's names), extracted from the non-structured texts. Document files, scanned from the Internet using the InfoStream system of content-monitoring, were used as an example [2].

While developing the network of concepts, algorithms of automated extraction of concepts from non-structured texts were used. Many works were devoted to these technologies (see, for example [3, 4]). It is worth mentioning that the approaches to the extraction of various types of concepts from the texts differ considerably both by their presentation context and structural features. To identify the document's belonging to a thematic column, requests made in a special way and in information-retrieval languages, including logic and context operators, parentheses, etc., can be used. To identify geographical names implies the use of tables, where, except for spelling templates of these names, country codes, region and town names are used. As an example, we can give a brief description of algorithm identification of company names in document texts. A document comes in the system entrance; it is analyzed in the process of sequent scanning. The document text is compared with templates, corresponding names of well-known firms, and in case of their existence, they are placed in a special table "document-firm". In addition, the extraction system of photographs envisages the identification of primarily unknown companies' names based on both templates and structural studying of the text. In particular, a table of suffix of the companies' names, containing such elements as "Inc", "Corp.", "Ltd", "Company" and others, is used.

Another kind of concepts, such as "persons", is extracted from texts based on the rules which consider tables of allowable names and surnames, initial templates, possible variants of joint spelling of initials/names and surnames.

It is important to state that the above-mentioned InfoStream system contains means of concept extraction, and presents them to the users in the form of "information portraits", which have such concepts as key words, geographical names, surnames of people, names of the companies etc.

The properties of the networks, formed with concepts, which are connected with each other by being mentioned in the same documents, are described in this work. The network formed with people's names, extracted from Internet-media text files according to common political topics during 1 (one) month and in 55 thousand documents, was studied more thoroughly. Over 19 thousand persons were mentioned in the texts.

As it has been found out in the framework of this research, the distribution of reference frequency of persons in the text file under consideration corresponds to Tsypfa law [3] (Fig. 1). The network of concepts, whose nodes are persons, and edges connecting the nodes correspond to the number of references of the persons in the same documents, is researched.

The network formed with concepts, extracted from text flows, is not static, and it depends on new documents which appear constantly, and corresponding concepts are extracted from them. Thus, to understand the structure of such network, it is necessary to take into account its evolution [6].

Let us look at some important characteristics from the theory of complex networks, which are considered in the context of this work.

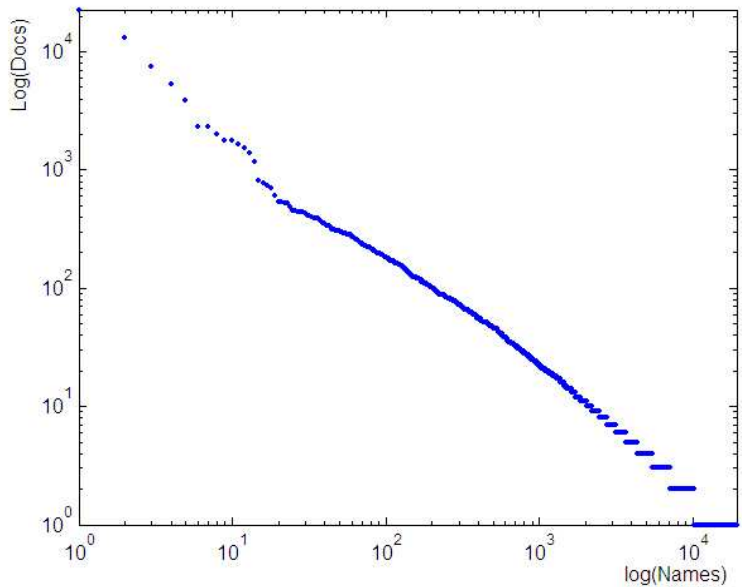

*Fig. 1. Plot of distribution of reference frequency of persons in a logarithmic scale*

The distance between nodes can be defined as a quantity of steps to be taken to get from one node to the other. Naturally, nodes can be connected directly or indirectly. It is possible to introduce a concept of an average distance for the whole system, i.e., the shortest way between pairs of nodes. But some networks can be unbound (a network of persons, for example), it means that there might be nodes with an infinite distance between them. Correspondingly, an average distance may appear to be infinite as well. To keep record of such cases, we introduce a concept of an average inverse distance between nodes, which is calculated as follows:

$$il = \frac{2}{n(n-1)} \sum_{i>j} \frac{1}{d_{ij}},$$

where $d_{ij}$ − the shortest distance between nodes $i$ and $j$.

The edges of the initial network are given weight meanings, equal to the number of documents (a document flow from Internet-media is analyzed), in which persons of corresponding nodes are mentioned. To prevent "noise", edges with the weight less than 2, were ignored. Developing the network with a fixed number of persons (e.g., in Fig. 2 a network with 50 persons is considered), which is realized through the increase of the number of documents under consideration, an average inverse distance between nodes increases reaching its logical saturation.

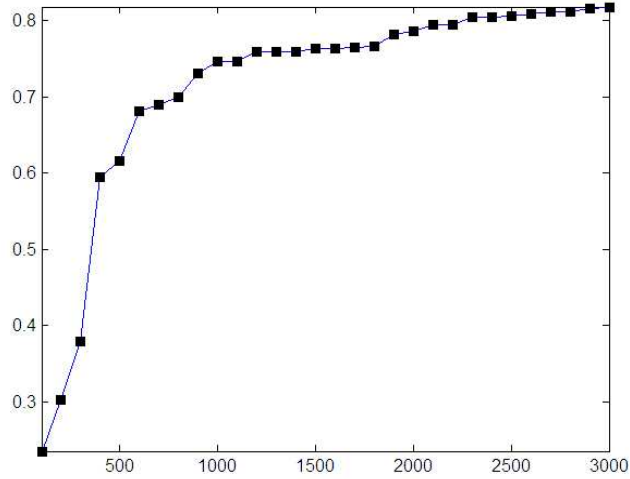

*Fig. 2. Dynamics of changing an average inverse distance (Y-axis)
when the number of documents increases (X-axis)*

The coefficient of clustering [6] characterizes the tendency to the development of groups of interconnected nodes, so-called cliques. For a separate node of the network, having a degree $k$, i.e., which $k$ edges come from, connecting it with other $k$ nodes (so-called the nearest neighbors), this parameter is defined as the ratio between a real quantity of edges, connecting the nearest neighbors among themselves, and a maximum possible one. If to assume that the nearest neighbors are connected directly with each other, the quantity of edges between them would be $\frac{1}{2}k(k-1)$. Hence, clustering coefficient is the number that corresponds to a maximum possible number of edges which could connect the nearest neighbors of a chosen node.

The level of clustering for the whole network is defined as a rated sum (based on the quantity of nodes) of corresponding coefficients of individual nodes. Naturally, when the network under consideration (consisting of a fixed quantity of concepts) is developed, and the number of analyzed documents increases, the quantity of edges increases constantly and clustering coefficient can reach meanings which are close to one (Fig. 3).

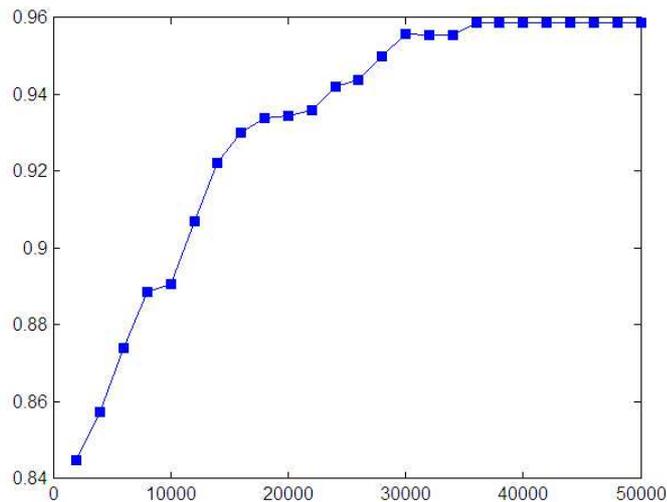

*Fig. 3. Dynamics of changing clustering coefficient (Y-axis)
when the number of documents increases (X-axis).*

One of the main characteristics of the network nodes is *betweenness*, which is similar to *load*, a term used in literary sources sometimes. This feature expresses the role of the node in establishing connections in the network and shows how many shortest ways come through it; it is also traditional for sociology where persons with a high level of *betweenness* play a leading role in establishing contacts with other persons. Obviously, betweenness coefficient ($b$) is complementary to clustering.

One of the results received in the context of this research is the establishment of the fact that nodes of the person network under consideration with a maximum quantity of edges (a degree) possess the highest level of betweenness in most cases (Fig.4); this is the reason why they can rather be viewed as the elements which connect separate person groups than as the basis for developing clusters under automated grouping.

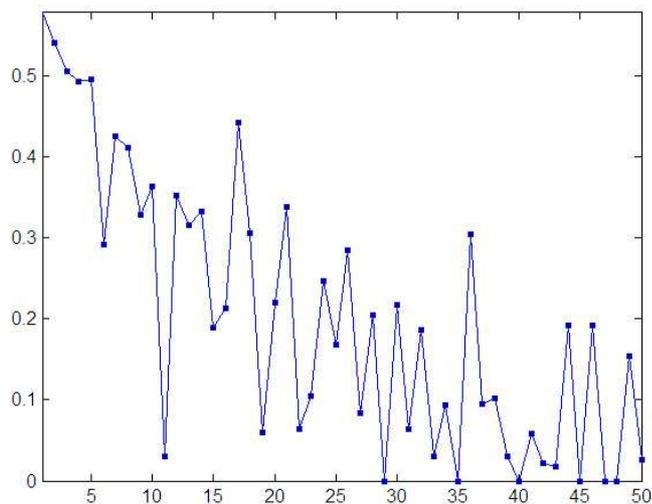

*Fig. 4. Coefficients $b$ (Y-axis) for nodes, ranged by a degree*

An important characteristic of the network is the distribution of node degrees $P(k)$, which is defined as probability that the node $i$ has a degree $k_i = k$. The networks, characterized by various $P(k)$, demonstrate different behavior. In some cases $P(k)$ can be Poisson distribution, exponential or degree distribution. The networks with exponential distribution of node degrees are called scale-free. It is scale-free distributions that are observed in really existing complex networks. The existence of nodes with a very high degree is possible in degree distribution; in fact they do not occur in the networks with Poisson or exponential distributions.

In a developing network under consideration with a fixed quantity of nodes corresponding to persons and an increasing number of documents, at first the distribution appeared to be close to a degree distribution and then to a Poisson distribution (Fig. 5). This is explained by the fact that at first node degrees have a systematic nature corresponding to real connections, and then due to a large quantity of "occasional contacts" which occur with large number of documents, the network becomes closer to an occasional one in which a great number of nodes are connected with numerous other ones (Fid. 6, 7).

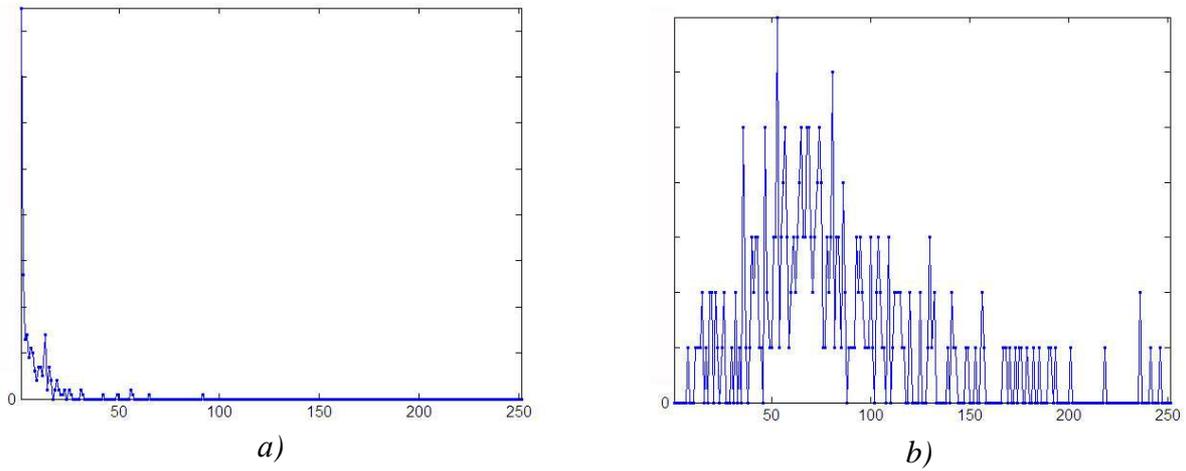

*Fig. 5. Distribution of network degrees:*
*a) Low relation of the scope of text files to the number of persons (1000:250);*
*b) high relation (50000:250)*

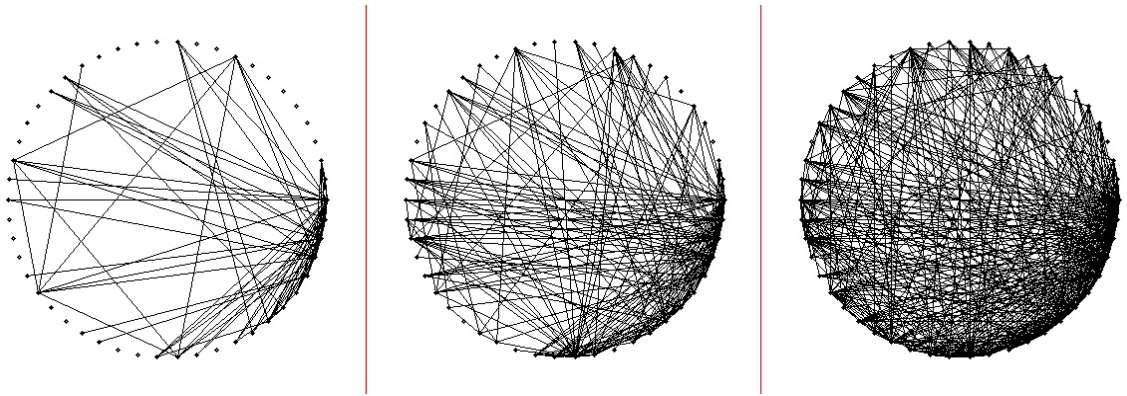

*Fig. 6. Dynamics of the network development when the number of documents in a text file increases*

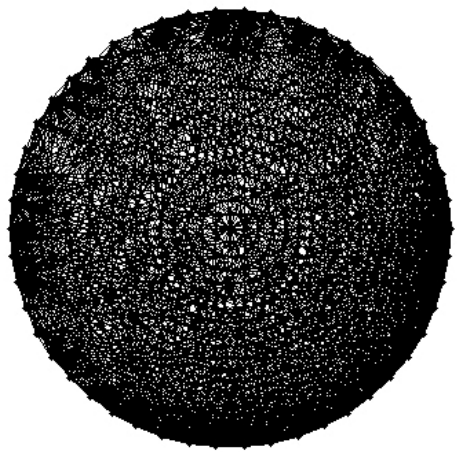

*Fig. 7. Network, close to a degradation condition: 50 persons, 50000 documents*

As it will be shown below, when the network under consideration is analyzed, the distribution of weight meanings of its edges, when various scopes of document flows are considered and whose rank distribution are shown in Fig. 8, is of great importance.

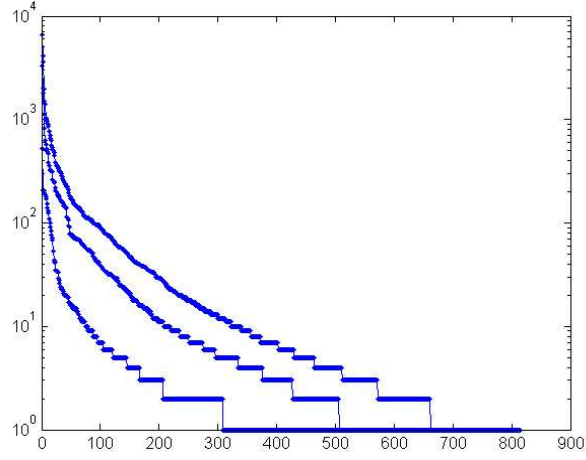

*Fig. 8. Distribution of network edge weight (X-axis) with 50 persons in a logarithmic scale (Y-axis) for web-publication files with 1000, 10000 and 50000 documents*

To avoid the network degradation, associated with the accumulation of "occasional contacts", let us determine a superimposed network, corresponding to the desired one, with some rough meanings of edge weights, namely, with help of the equation:

$$v' = \begin{cases} 1, & v \geq \varepsilon v_{max} \\ 0, & v < \varepsilon v_{max} \end{cases}$$

where $v'$ - weight of the edge of a superimposed network, $v$ - weight of the edge of the initial person network, $v_{max}$ - maximum meaning of the edge weight, $\varepsilon$ - coefficient of rough estimate.

As the measurements show, weight meanings of the network edges are distributed exponentially on a larger area; it allows assuming:

$$v_i = e^{a_i + \lambda r},$$

where $v_i$ - edge weight of a person network, corresponding to a certain number of input documents $D_i$, $a_i$ - a coefficient, depending on meaning $D_i$, $\lambda$ - a constant, $r$ - meaning of the edge rank (numbers of decreasing ranking of weight meanings of edges).

Let us assume that for some $r = r_\varepsilon$ ($0 < \varepsilon \leq 1$) the following is performed:

$$v_i = e^{a_i + \lambda r_\varepsilon} = \varepsilon e^{a_i}.$$

In this case for some quantity of input documents $D_k$ for the same meaning $r = r_\varepsilon$ the following will be performed:

$$v_k = e^{a_k + \lambda r_\varepsilon} = e^{a_k - a_i + a_i + \lambda r_\varepsilon} = \varepsilon e^{a_i} \times e^{a_k} \times e^{-a_i} = \varepsilon e^{a_k}.$$

Thus, in accordance with a suggested model, where $\varepsilon$ expresses a threshold meaning in the conditions of rough estimate of a superimposed model, total meanings of all $v'$ (a quantity of edges in a superimposed network) appear to be a constant quantity.

The studying of real data showed that the meanings of an average distance and clustering appeared to be constant as well. The effect proves the stability of a superimposed network and its relative independence from the scopes of in-coming documents. In particular, for meaning $\varepsilon = 0.001$, 50 persons and the quantity of documents ranging from 1000 to 50000, a clustering coefficient was $0.78 \pm 0.01$, and the average inverse distance - $0.65 \pm 0.02$.

The empiric results received can be useful, for example, for theoretical description and modeling of social and technological processes, identifying and visualizing implicit connections of separate objects or subjects in a competitive survey.

The **stabilization phenomenon of a superimposed network** makes it possible to practically identify stable connections, to reduce the effect of noise factors through the analysis of relatively small document files. Alongside with this, the issue of the estimation of the correlation of the received information person connections calculated by counting document frequency, where persons are mentioned together and real interconnections, remains open/not studied.

It is necessary to state that the stabilization of a superimposed network was studied where $\varepsilon$ was higher than a threshold unit. At the same time, in view of an exponential nature of the distribution of weight meanings of the edges in the initial network, probably the part of connections, ignored by us, is as complicated as the whole network.

In a conclusion, the authors would like to thank the staff members of the Information center ElVisti, S. M. Braichevskiy and A.T. Darmokhval for their participation in constructive discussions of particular aspects, presented in this work, and for their assistance in making calculations.